\documentclass[aps,prd,showpacs,preprintnumbers,nofootinbib,amsmath,amssymb]{revtex4}

\usepackage{graphicx}
\usepackage{dcolumn}
\usepackage{bm}
\usepackage{amsmath}

\newcommand{\beq}{\begin{eqnarray}}
\newcommand{\eeq}{\end{eqnarray}}

\newcommand{\real}{{\sf I}\kern-.12em{\sf R}}
\newcommand{\comp}{{\sf I}\kern-.50em{\sf C}}
\newcommand{\unity}{{\sf I}\kern-.54em{\sf 1}}

\newcommand{\rep}[1]{(\ref{#1})}

\def\spose#1{\hbox to 0pt{#1\hss}}
\def\ltapprox{\mathrel{\spose{\lower 3pt\hbox{$\mathchar"218$}}
 \raise 2.0pt\hbox{$\mathchar"13C$}}}

\begin{document}

\title{Thermal Monopole Condensation and Confinement\\ in finite temperature 
Yang-Mills Theories}
\author{Alessio D'Alessandro, Massimo D'Elia$^1$ and Edward V. Shuryak$^2$} 
\affiliation{$^1$Dipartimento di Fisica, Universit\`a
di Genova and INFN, Via Dodecaneso 33, 16146 Genova, Italy\\
$^2$Department of Physics and Astronomy, 
State University of New York, 
Stony Brook NY 11794-3800, USA}
\date{\today}

\begin{abstract}
We investigate the connection between Color Confinement and
thermal Abelian monopoles populating
the deconfined phase of SU(2) Yang-Mills theory, by studying how the 
statistical properties of the monopole ensemble change as the 
confinement/deconfinement temperature is approached from above.
In particular we study the distribution of monopole currents with multiple
wrappings in the Euclidean time direction, corresponding to two or more particle
permutations, and show that multiple wrappings increase as the deconfinement
temperature is approached from above, in a way compatible with a condensation
of such objects happening right at the deconfining transition.
We also address the question of the thermal monopole mass, showing 
that different definitions give consistent results only around
the transition, where the monopole mass goes down 
and becomes of the order of the critical temperature itself. 
\end{abstract}

\maketitle

\section{Introduction}

Color confinement is not yet fully understood in terms of the first principles
of Quantum Chromodynamics (QCD). Models exist which relate confinement to the condensation
of topological defects in the QCD ground state; one of them
is based on dual superconductivity of the QCD
vacuum~\cite{thooft75,mandelstam}.
According to this model, color confinement is due to the
spontaneous breaking of a magnetic simmetry, induced by the condensation
of magnetically charged defects (e.g. magnetic monopoles), which yields a
non-vanishing magnetically charged Higgs condensate.

The existence of a new phase of matter, in which quark and gluons are deconfined 
(Quark-Gluon Plasma), is a well defined prediction of lattice QCD simulations: the 
deconfined phase is under active experimental search in heavy ion experiments. 
However the physical properties expected for this phase are not yet clearly
understood: the Quark-Gluon Plasma (QGP) 
is still strongly interacting above the 
deconfining temperature $T_c$ and its properties may be more similar to those
of an almost perfect liquid.

One of the hypotheses which have been put forward in the recent past is that QGP properties may be 
dominated by a magnetic component~\cite{shuryak,chezak,Shuryak:2008eq}. 
In Ref.~\cite{chezak} such magnetic
component has been related
to thermal Abelian monopoles evaporating from the magnetic condensate which is 
believed to induce color confinement at low temperatures;
moreover it has been proposed to detect such thermal monopoles in 
finite temperature lattice QCD simulations, by identifying them 
with monopole currents having a non-trivial wrapping in the Euclidean 
temporal direction~\cite{chezak,bornya92,ejiri}. 
First numerical investigations
of these wrapping trajectories were performed in Ref.~\cite{bornya92}
and~\cite{ejiri}, while a systematic study, regarding 
the deconfined phase of $SU(2)$ Yang-Mills theory, has been performed
in Ref.~\cite{monden}.

The definition of Abelian magnetic monopoles in non-Abelian gauge theories requires 
the identification of Abelian degrees of freedom: that is done usually by a 
procedure known as Abelian projection and relies on the choice of an adjoint
field. Since no natural adjoint field exists in usual QCD, that implies some
arbitrariness; a popular choice is to perform the projection in 
the so-called Maximal Abelian gauge (MAG).
Results obtained in Ref.~\cite{monden} have shown that, as already well known for 
Abelian monopole currents in general, also the number and the locations
of monopole currents with a non-trivial wrapping in the Euclidean time direction
are quantities which depend on the choice of the Abelian projection. 

Despite that,
the density and the spatial correlation functions of MAG thermal monopoles
show a negligible dependence 
on the UV cut-off~\cite{monden}, as expected for a physical quantity.
The temperature dependence of the monopole density,
$\rho$, is not compatible with a (massive or massless) free
particle behavior and is instead well described,
in the whole range of temperatures explored,
by a behavior
$\rho \propto T^3/(\log T/\Lambda_{eff})^2$ with
$\Lambda_{eff} \sim 100$ MeV, while the behavior 
$\rho \propto T^3/(\log T)^3$, predicted 
by dimensional reduction arguments, is compatible with 
data for  $T > 5\ T_c$. 
This is in agreement with the picture of an electric dominated
phase for Yang--Mills theories at very high temperatures, in which 
the magnetic component is strongly interacting~\cite{shuryak}. 

Moreover the study of density--density spatial correlation functions 
has verified the presence of 
a repulsive (attractive) interaction for a monopole--monopole (monopole--antimonopole) pair,
which at large distances is in agreement with a screened Coulomb potential and a 
screening length of the order of 0.1 fm. The above results have suggested a liquid-like behavior for the 
thermal monopole ensemble above $T_c$~\cite{liquid} and stimulated further research
about the possible role of magnetic monopoles in the 
Quark-Gluon Plasma~\cite{ratti,dyons,lublin}.

In the present paper, while aware of the problems related to the definition of 
thermal Abelian monopoles, we work on the hypothesis that those
defined in the MAG may have a physical meaning and address a question
regarding color confinement:
if thermal monopoles in the deconfined 
phase are really related to the magnetic condensate responsible for
confinement below the deconfinement temperature, $T < T_c$,
 is it possible to find clear signatures for 
their approach to condensation for temperatures slightly above $T_c$? Such
approach would complement standard studies about the validity of the dual
superconductor model, which look at the spontaneous 
breaking of a magnetic symmetry in the confined phase (i.e. the presence
of a magnetic condensate) and at its restoration as $T_c$ is approached 
from below~~\cite{superI-II,superIII,superfull,superIV,moscow,bari,vacuumtype,conradi}.

Since condensation is a phenomenon strictly related to quantum statistics,
any signal of it should be searched for in properties of thermal
Abelian monopole trajectories which reflect their nature of identical quantum objects.
Such properties, as better explained in the following, are 
encoded in monopole trajectories which wrap two or more times in the Euclidean time
direction and correspond to the permutation of two or more particles: no or very few multiple wrapping
trajectories are expected if the system is very close to the Boltzmann approximation,
while their number should increase as quantum effects becomes important, in a critical 
way close to the point where the particles condense.
Such approach goes
back to the seminal papers by Feynman~\cite{fey1,fey2}, where 
a path integral formulation was applied to 
describe condensation phenomena like
the superfluid transition in $^4$He. Since then, the path
integral formulation of multi-particle systems has become an invaluable tool for 
analytical and numerical investigations of Bose condensation~\cite{elser,ceperley},
and has been recently reconsidered in Ref.~\cite{cristoforetti}, where it has been suggested
to apply it also to the analysis of monopole condensation in QCD.

A second question that we address regards the mass of thermal monopoles. 
Different strategies can be followed to determine some temperature
dependent effective mass. For instance, one can 
look at the effective mass appearing in the description of multiply
wrapping trajectories above $T_c$ or,
in close analogy
with the path integral formulation for non-relativistic quantum particles, 
one can obtain information about
the mass by looking at the spatial fluctuations of wrapping trajectories.
Of course it is not guaranteed that different strategies will lead to consistent 
determinations.

The paper is organized as follows. In Section~\ref{section2} we recall a few technical details
about the definition of thermal Abelian monopoles on the lattice. In Section~\ref{section3} 
we study
the distribution
of multiply wrapping monopole trajectories in the deconfined phase of the SU(2) pure gauge theory 
and give an interpretation of them
in terms of monopole condensation, determing a condensation temperature
which is consistent with the known deconfinement temperature $T_c$.  
In Section~\ref{section4} 
we address the question of the monopole mass.
Finally, in Section~\ref{section5}, we discuss our results
and draw our conclusions.

\section{Abelian Projection and Thermal Monopoles on the lattice}
\label{section2}

The definition of Abelian monopoles requires the identification of 
a U(1) subgroup of the original gauge group. 
This is done by a procedure known as Abelian projection, which 
is assigned in terms of an adjoint field, i.e., in  the particular
case of $SU(2)$ color gauge group, 
a vector field $\vec \phi (x)$ transforming as
\beq \vec \sigma \cdot \vec \phi (x) \to 
G(x) (\vec \sigma \cdot \vec \phi (x)) G^\dagger (x)
\eeq
under a local gauge transformation $G(x)$ 
($\vec \sigma$ denote Pauli matrixes). If $\vec G_{\mu\nu}(x)$ is the field
strength tensor of the original SU(2) gauge theory, then an Abelian
tensor, known as 't Hooft
tensor, is defined in terms of $\vec \phi (x)$
\beq
 F_{\mu\nu} &=& \hat\phi\cdot\vec G_{\mu\nu} - \frac{1}{g}
\hat\phi\cdot(D_\mu \hat\phi\wedge D_\nu \hat\phi) \nonumber \\
&=& \partial_\mu(\hat\phi\cdot \vec A_\nu) -
\partial_\nu(\hat\phi\cdot \vec A_\mu) -\frac{1}{g}
\hat\phi\cdot(\partial_\mu\hat\phi\wedge\partial_\nu\hat\phi)
\eeq
where $\hat \phi(x) \equiv \vec \phi(x) / |\vec \phi(x)|$. 
$F_{\mu\nu}$ is an Abelian, gauge independent 
tensor, which 
in the gauge where 
$\hat\phi (x) = (0,0,1)$ takes the form
\beq
F_{\mu\nu} = \partial_\mu A^3_\nu -
\partial_\nu A^3_\mu \nonumber \, .
\eeq
In that gauge, 
which is fixed up to a $U(1)$ residual gauge freedom
($\hat \phi \in SO(3)/U(1)$),
the Abelian projection
corresponds to taking the diagonal part of gauge links.

In usual QCD there is no natural adjoint field, but several adjoint
fields can be constructed in terms of gauge fields, typically
a closed parallel transport (i.e. a path ordered product of gauge links).
Alternatively, an implicit definition can be assigned by fixing 
$\hat\phi = (0,0,1)$ and constant in one particular gauge: that
defines $\hat\phi (x)$ in every other gauge. An example is the 
so-called Maximal Abelian Gauge, where the gauge is fixed by
maximizing the following functional with respect to gauge transformations:
\begin{equation}
F_{\rm MAG} = \sum_{\mu,n} {\rm Re}\,  \mbox{tr} \left[U_\mu(n)
  \sigma_3 U^{\dagger}_\mu(n) \, 
\sigma_3\right]
\label{maxfun}
\end{equation}
where $U_\mu(n)$ is a non-Abelian gauge link variable, i.e. an
elementary parallel transport from lattice site $n$ 
to $n + \hat\mu$.
$F_{\rm MAG}$ is proportional to the average squared diagonal part of the
gauge links.

In the gauge where the 't Hooft tensor is diagonal, Abelian link phases
are extracted as follows:
\beq
U_\mu(n) = u_\mu^0 {\rm Id} + i \vec \sigma \cdot \vec u_\mu \to 
{\rm diag} (e^{i \theta_\mu(n)},e^{-i \theta_\mu(n)}) \equiv 
\frac{(u_\mu^0 {\rm Id} + i \sigma_3 u_\mu^3)}{
\sqrt{(u_\mu^0)^2 + (u_\mu^3)^2}} \, .
\eeq
Abelian plaquettes, i.e. the lattice discretization of the
't Hooft field strength tensor, can then be constructed starting
from the Abelian gauge link phases $\theta_\mu(n)$:
\beq \theta_{\mu\nu} \equiv 
\theta_\mu(n) + \theta_\nu(n + \hat\mu) - \theta_\mu(n + \hat\nu)
- \theta_\nu(n) \equiv
\hat\partial_\mu \theta_\nu - \hat\partial_\nu \theta_\mu \, .
\eeq
Of course Abelian plaquettes can be extracted also in a gauge 
where $\hat\phi(x)$ is not constant, even if through a more
intricate procedure~\cite{DTkondo},
obtaining equivalent results~\cite{DTkondo,monden_proc}, as expected
from the gauge invariance of 't~Hooft tensor.

Abelian monopole currents are identified by
the standard De Grand-Toussaint construction~\cite{degrand}:
\begin{equation}
m_\mu = {1 \over 2 \pi} \varepsilon_{\mu\nu\rho\sigma} \hat\partial_\nu \overline
\theta_{\rho\sigma}
\end{equation}
where $\overline \theta_{\rho\sigma}$ is the compactified part of the
Abelian plaquette phase
\begin{equation}
\theta_{\mu\nu}= \overline \theta_{\mu\nu} + 2 \pi n_{\mu\nu}
\end{equation}
and $n_{\mu\nu}\in \mathbf{N}$.

Monopole currents form closed loops,
since $\hat\partial_\mu m_\mu=0$. These loops may be either
topologically trivial or wrapped around the lattice.
Following the proposal of Ref.~\cite{chezak}, 
among the different currents we select those having a non-trivial 
positive (negative) wrapping around the Euclidean temporal direction, 
identifying them
with thermal monopoles (antimonopoles).  

This procedure
identifies, at any given timeslice, the spatial positions
of thermal monopoles with the points pierced by a current with non-trivial
wrapping. There is some sort of ambiguity in this identification, since 
monopole currents extracted on the lattice are not strictly one-dimensional 
objects, but rather clusters of currents becoming approximately one-dimensional
tubes in the high temperature phase, but staying anyway of finite thickness in 
lattice spacing units, because of ultraviolet (UV) noise which
is present in the form 
of small monopole loops attached to the wrapping current.
 In our numerical
setup we fix the spatial position of the monopole at the point where
the wrapping monopole current is first detected by the current 
searching algorithm: thermal monopoles are thus spatially located
with some random noise, which however has negligible systematic effects
on the density and spatial correlation functions of thermal monopoles.

Some of the monopole currents may wrap more than once,
say $k$ times, in the Euclidean time direction: in this case,
as better explained in the following, we interpret them 
as sets of $k$ monopoles (or antimonopoles) undergoing a cyclic
permutation after a wrap in Euclidean time. The density and distribution
of those $k$-times wrapping trajectories is one of the subject 
of the present investigation. Also in this case, mainly due to 
the finite thickness of the wrapping current induced by UV fluctuations, 
ambiguities may be present for some specific configurations. 
One can imagine of situations in which it is hard to distinguish
between two currents, each wrapping once, and a single current with 
a double wrap: that may happen, for instance, if the two currents
overlap in some point and the answer may depend on the algorithm
used to follow the current around the lattice. 
In our measurements we have 
chosen different current searching 
algorithms and we have taken the discrepancies between the different
algorithms as an estimate of the systematic error linked to such
ambiguities: such systematic error is always included in our 
determinations.

In the following we shall present results obtained in the 
Maximal Abelian Gauge.
It can be shown that, on stationary points of the MAG
functional, the local operator
\begin{equation}
X(n) = \sum_\mu \left[ U_\mu(n) \sigma_3 U^\dagger_\mu(n)
+U^\dagger_\mu(n-\hat\mu) \sigma_3 U_\mu(n-\hat\mu)\right]
\end{equation}
is diagonal.
The maximization of the MAG functional for a given
configuration has been achieved by an iterative combination of local 
maximization and overrelaxation (see Ref.~\cite{cosmai}), 
stopping the algorithm
when the average squared modulus of the non-diagonal part
of $X(n)$ was less than a given parameter $\omega$~\cite{cosmai}. 
We have chosen $\omega = 10^{-8}$
(see Ref.~\cite{monden} for more details).

\section{Monopole condensation}
\label{section3}

If we interpret the set of 
wrapping monopole trajectories, extracted from one gauge configuration of our Monte-Carlo
sample, as one possible configuration of the Euclidean path integral representation 
of an 
ensemble of identical monopoles and antimonopoles at thermal equilibrium, 
then a trajectory
which wraps $k$ times before closing can be interpreted as a set of $k$ monopoles (or antimonopoles)
which permutate cyclically after going through the periodic Euclidean time direction.

\begin{figure}
\begin{center}
\includegraphics*[width=0.8\textwidth]{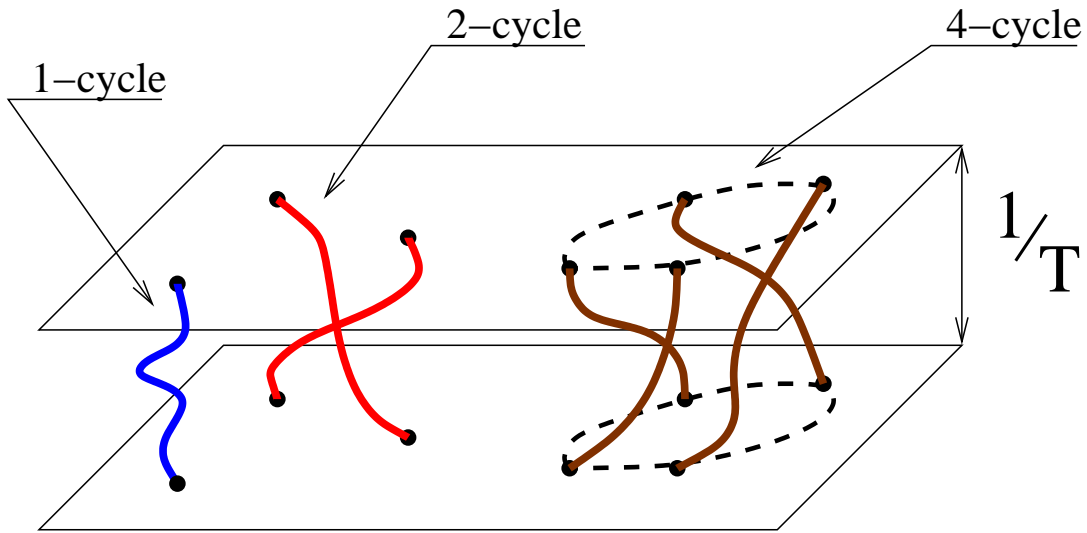}
\caption{
A possible contribution to the path integral representation of the 
partition function of 7 identical particles, in which the particles
undergo a permutation made up of a 1-cycle, a 2-cycle and a 4-cycle.
}
\label{fig1}
\vspace{1cm}
\end{center}
\end{figure}

In the path integral describing $N$ 
identical particles at thermal equilibrium, 
each possible configuration of the $N$ particle paths contributing to the functional integral
needs to be periodic, apart from a possible permutation of the $N$ particles (the sign of the permutation
is attached to the contribution if the particles are fermions). That means 
that the configuration is not necessarily composed of $N$ closed paths (that would correspond
to the identical permutation), but is in general made up of $M$ closed paths, with $M \leq N$:
if the $j$-th path wraps $k_j$ times around the Euclidean time direction then 
$\sum_{j = 1}^M k_j = N$ and the configuration corresponds to a permutation made up 
of $M$ cycles of sizes $k_1,k_2,\ \dots\ k_M $. In Fig.~\ref{fig1} we report
an example corresponding to a permutation of 7 particles partitioned into
a 1-cycle, a 2-cycle and a 4-cycle.

If effects related to quantum statistics are negligible, i.e. if the system is very close to 
the Boltzmann approximation, configurations corresponding to permutations different from 
the identical one are expected to have a negligible weight in the path integral, so that
trajectories wrapping more than one time, corresponding to the exchange of two or more
particles, are very rare. The number of trajectories wrapping more and more times is 
instead expected to increase as quantum effects become more important, and this should be
especially true close to a transition associated with
Bose-Einstein Condensation (BEC) (or with similar phenomena).

\subsection{Numerical simulations}

In Table~\ref{tabwrap} we report the normalized average densities $\rho_k/T^3$ 
of trajectories wrapping $k$ times as a function of temperature, determined 
for SU(2) pure gauge theory by use of the standard plaquette action 
and of the MAG Abelian projection.
Data include both monopole and antimonopole trajectories and have 
been obtained by extracting
monopole currents from samples consisting of $O(10^3)$ independent 
gauge configurations
for each value of $T$. The superscript $a$ or $b$ above the temperature 
value refers
to two different values of the inverse gauge coupling, $\beta = 2.70$ and 
$\beta = 2.60$, corresponding respectively to lattice spacings 
$a(\beta) \simeq 0.047$ fm and $a(\beta) \simeq 0.063$ fm\footnote{
The physical scale
has been determined according to
$a(\beta) \Lambda_L = R(\beta) \lambda(\beta)$, where $R$ is the 
two-loop perturbative {$\beta$-function}, while $\lambda$ is a 
non-perturbative correction factor computed and reported in Ref.~\cite{karsch}.
We have assumed the values $T_c / \Lambda_L = 21.45(14)$~\cite{karsch}, 
$T_c / \sqrt{\sigma} = 0.69(2)$~\cite{karsch2} and
$\sqrt{\sigma} \simeq 430$ MeV.
}. Simulations have been done at fixed spatial volume, while the temperature
$T = 1/(N_t a (\beta))$ has been changed by varying the number of lattice
sites $N_t$ in the temporal direction. In particular, 
at $\beta = 2.70$ we have made simulations on
$64^3 \times N_t$ lattices, with $N_t = 4,5,\dots,14$, while 
at $\beta = 2.60$ we have chosen 
$48^3 \times N_t$ lattices, with $N_t = 4,5,\dots,10$.
Different spatial volumes have been considered in some cases for 
$\beta = 2.60$, in order to check for finite size effects.
Confirming results reported for the overall monopole-antimonopole density 
in Ref.~\cite{monden}, data obtained for each $\rho_k$ show a reasonable independence
from the value of the UV cutoff, even if small deviations are visible.

\begin{figure}
\begin{center}
\includegraphics*[width=0.7\textwidth]{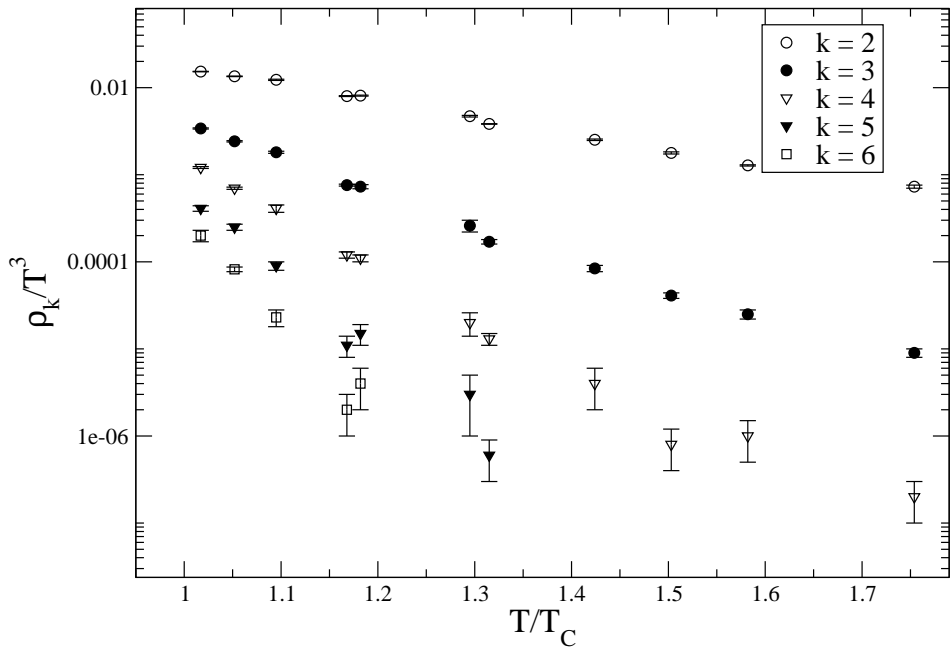}
\caption{
Normalized densities $\rho_k/T^3$ as a function of $T/T_c$.
}
\label{rhok_t}
\vspace{1cm}
\end{center}
\end{figure}

Numerical data for $\rho_k/T^3$
are also reported in Fig.~\ref{rhok_t}.
What is apparent from the figure is that the relative weight of trajectories wrapping more 
than one time increases rapidly as $T$ approaches $T_c$ from above. The number of monopoles
or antimonopoles which are found in non-trivial cycles (i.e. $k > 1$) is 
less than 0.1\% for $T > 2.5\, T_c$, meaning that the system is essentially
Boltzmann-like at high temperatures. The same number goes to about 1\% for $T \sim 1.5 T_c$
and is well above 10\% at the lowest temperature explored in our simulations.

This is a clear qualitative signature for quantum statistics effects becoming more and more relevant
as the critical temperature is approached, as expected if monopoles condense at $T_c$.
In the following we shall try to be more quantitative, but in order to do that we need to rely on some 
specific model. 
We shall first consider the analysis of a system of non-relativistic non-interacting bosons, 
following what performed in Refs.~\cite{fey2,elser} as a starting point 
to describe the transition to superfluid helium.
We shall then discuss the deviations from this very simple model which are expected 
for the monopole-antimonopole ensemble in QCD.

\begin{table}
\begin{center}
\begin{tabular}{|c|c|c|c|c|c|c|c|c|}
\hline
$T/T_c$  &  $\rho_1/T^3$  &  $\rho_2/T^3$  &  $\rho_3/T^3$  &  $\rho_4/T^3$  &  $\rho_5/T^3$  &  $\rho_6/T^3$  &  
$\rho_7/T^3$  &  $\rho_8/T^3$  
\\
\hline $1.017^a$ & 0.308(2) & $1.53(1)\ 10^{-2}$ & $3.40(5)\  10^{-3}$ & $1.21(3)\ 10^{-3}$ & $4.1(3)\ 10^{-4}$ & $2.0(3)\ 10^{-4}$ &
$0.6(2)\ 10^{-4}$ & $2.4(5)\ 10^{-5}$\\
\hline $1.052^b$ & 0.315(5)& $1.35(1)\ 10^{-2}$& $2.42(4)\  10^{-3}$ & $7.0(2)\ 10^{-4}$  & $2.5(2)\ 10^{-4}$ & $8.2(5)\ 10^{-5}$ 
& $3.6(5)\ 10^{-5}$ & $1.1(3)\ 10^{-5}$ \\
\hline $1.095^a$ & 0.3395(15) & $1.23(2)\ 10^{-2}$  & $1.81(5)\ 10^{-3}$ & $4.1(4)\ 10^{-4}$  & $0.9(1)\ 10^{-4}$ & $2.3(5)\ 10^{-5}$ &
$1.0(5)\ 10^{-5}$ & \\
\hline $1.168^b$ & 0.325(3) & $8.0(1)\ 10^{-3}$ & $7.6(2)\ 10^{-4}$  & $ 1.2(1)\ 10^{-4}$ & $1.1(3)\ 10^{-5}$ & $0.2(1)\ 10^{-5}$&&\\
\hline $1.187^a$ & 0.337(2) & $8.1(1)\ 10^{-3}$  & $7.3(4)\ 10^{-4}$  & $ 1.1(1)\ 10^{-4}$  & $1.5(4)\ 10^{-5}$ & $0.4(2)\ 10^{-5}$& &\\
\hline $1.295^a$ & 0.316(1) & $4.72(10)\ 10^{-3}$  & $2.6(3)\ 10^{-4}$  & $2.0(6)\ 10^{-5}$  & $0.3(2)\ 10^{-5}$ & & & \\
\hline $1.315^b$ & 0.297(2) & $3.83(3)\ 10^{-3}$ & $1.7(1)\ 10^{-4}$  & $1.3(2)\ 10^{-5}$  & $0.6(3)\ 10^{-6}$ & & & \\
\hline $1.424^a$ & 0.286(1) & $2.52(5)\ 10^{-3}$ & $8.4(7)\ 10^{-5}$  & $0.4(2)\ 10^{-5}$ & & & &\\
\hline $1.503^b$ & 0.271(1) & $1.78(5)\ 10^{-3}$ & $4.1(3)\ 10^{-5}$ & $0.8(4)\ 10^{-6}$  & & & &\\
\hline $1.582^a$ & 0.252(1) & $1.28(2)\ 10^{-3}$ & $2.5(3)\ 10^{-5}$ & $1.0(5)\ 10^{-6}$  & & & &\\
\hline $1.754^b$ & 0.2134(10) & $7.3(3)\ 10^{-4}$ & $9(1)\ 10^{-6}$ & $0.2(1)\ 10^{-6}$  & & & &\\
\hline $1.780^a$ & 0.2190(2) & $6.26(7)\ 10^{-4}$ & $8.3(8)\ 10^{-6}$ & $0.2(1)\ 10^{-6}$  & & & &\\
\hline $2.034^a$ & 0.1870(4) & $3.04(10)\ 10^{-4}$ & $1.0(4)\ 10^{-6}$ & & & & &\\
\hline $2.105^b$ & 0.1745(10) & $3.02(20)\ 10^{-4}$ & $1.7(4)\ 10^{-6}$ & & & & &\\
\hline $2.373^a$ & 0.1574(4) & $1.30(6)\ 10^{-4}$ & $0.4(2)\ 10^{-6}$ & & & & & \\
\hline $2.631^b$ & 0.1410(10) & $1.30(10)\ 10^{-4}$ & $0.3(1)\ 10^{-6}$ & & & & &\\
\hline $2.848^a$ & 0.1306(4) & $6.3(2)\ 10^{-5}$  & $0.2(1)\ 10^{-6}$ & & & & &\\
\hline $3.560^a$ & 0.1073(3) & $3.3(4)\ 10^{-5}$  & $0.4(4)\ 10^{-7}$ & & & & &\\
\hline
\end{tabular}
\end{center}
\caption{Normalized average densities $\rho_k/T^3$ 
of trajectories wrapping $k$ times, as a function of $T/T_c$  
for SU(2) pure gauge theory. The superscript $a$ or $b$ above the temperature 
value refers
to two different values of the lattice spacing, $0.047$ fm and 
$0.063$ fm respectively. Blank spaces indicate that no monopole current
with the given wrapping has been observed in our statistical sample.
}
\label{tabwrap}
\end{table}

\subsection{Non-interacting Model for BEC}

\label{freebec}

We consider a system of $N$ identical bosonic particles and their partition function 
$Z = {\rm Tr} (e^{-H/T})$. 
The trace can be taken over position eigenstates $|x_1 \ldots x_N \rangle$,
but the correct states to consider in case of identical particles are proportional\footnote{
The normalization factor is $1/\sqrt{N!}$ if the $N$ coordinates are all different and changes otherwise.} to
$$ \sum_P | x_{P_1} \ldots x_{P_N} \rangle$$
where the sum is over all the possible permutations $P$ of the $N$
particles. The partition function can be written as
\beq
Z=\frac{1}{{N!}} \sum_P \int d^3x_1 \ldots \int d^3x_N \langle x_{P_1} \ldots x_{P_N} | e^{-\beta H} |x_1 \ldots x_N\rangle \, .
\label{Zbec}
\eeq
We keep only the kinetic term $K = p^2/(2m)$ of the Hamiltionian $H$, discarding
interactions and relativistic effects:
we shall discuss these approximations in Section~\ref{monbec}.

The assumption of absence of particle-particle interactions implies that the matrix element in Eq.~(\ref{Zbec})
can be conveniently factorized according to the decomposition of the permutation into disjoint
cycles. To clarify this point, consider the following explicit case involving 5 particles
and a permutation composed of a 3-cycle plus a 2-cycle:
\beq
&& \langle x_3, x_1, x_2, x_5, x_4 | e^{-H/T} |x_1, x_2, x_3, x_4, x_5 \rangle = \nonumber \\ 
&& \langle x_3, x_1, x_2 | e^{-(p_1^2 + p_2^2 + p_3^2)/(2mT)} |x_1, x_2, x_3 \rangle  
\langle x_5, x_4 | e^{-(p_4^2 + p_5^2)/(2mT)} |x_4, x_5 \rangle \, . 
\eeq

Also the integration in Eq.~(\ref{Zbec}) can be carried on independently over groups of $x$ variables
belonging to the same cycle, so that each summand permutation can be factorized into a product of different
contributions and the partition function can be rewritten as follows:
\beq
Z=\frac{1}{{N!}} \sum_P \prod_k z_k^{n_k}
\label{Zbec2}
\eeq
where 
\beq
z_k \equiv
\int d^3y_1 \ldots \int d^3y_k\ \langle y_2, y_3 \ldots y_{k}, y_1 | e^{-(p_1^2 + \dots  + p_k^2)/(2 m T)} |y_1, y_2 \ldots y_k\rangle
\label{kcycle}
\eeq
is the contribution coming from one $k$-cycle and $n_k$ is the number of different $k$-cycles appearing
in the cycle decomposition of the permutation $P$. It is straightforward to see that $z_k$ can be rewritten as
\beq
z_k(T) &=& \int d^3y_1 \ldots \int d^3y_k    \langle y_1 | e^{-{p_k^2\over 2mT}} |y_k \rangle \dots \langle y_3 | e^{-{p_2^2\over 2mT}} |y_2 \rangle 
\langle y_2 | e^{-{p_1^2 \over 2mT}} |y_1 \rangle \nonumber \\
&=& \int d^3y_1 \langle y_1 | e^{- k {p^2 \over 2mT}} |y_1 \rangle = z_1(T/k)
\label{kcycle2}
\eeq
where $z_1(T/k)$ is the partition function for a single particle at a temperature
$k$ times lower. Notice that the whole derivation would be unchanged if the particles move in an external
potential, but still in absence of particle-particle interactions.

The partition function for a non-relativistic free particle in a box is well known:
\beq
z_1(T) = {V}/{\lambda^3}
\label{1pz}
\eeq
where $V$ is the available volume and 
$\lambda=\sqrt{2\pi/(m T)}$ is the De Broglie thermal wavelength. There is an approximation involved in the result
above: the spacing between the quantum energy levels of the particle must be negligible with respect
to $T$, so that the sum over energy levels can be replaced by an integral. That condition can be written
equivalently, for a cubic box, as $\lambda \ll V^{1/3}$, i.e. the thermal wavelength must be neglibible
with respect to the size of the box; a similar condition applies if the particle is constrained in 
a three-dimensional periodic torus, as happens in our lattice setup. 
For a given volume, this condition is violated at low enough temperatures.

Combining Eqs.~(\ref{kcycle2}) and (\ref{1pz}) we obtain
\beq
z_k(T) = z_1(T/k) = \frac{V}{\lambda^3 k^{3/2}}
\label{kcycle3}
\eeq
but the condition for its validity is now $\sqrt{k}\ \lambda \ll V^{1/3}$. Such condition can be violated, for given 
$V$ and $T$, if $k$ is large enough. To better understand this condition, 
think of the loop
constructed with the positions of the $k$ particles belonging to the cycle; 
the integral in Eq.~(\ref{kcycle}) can be rewritten in terms of integration 
over the relative coordinates of neighbooring particles plus 
an integration over the center of mass of the loop: only if the size of the loop,
which is of order $\sqrt{k}\ \lambda$, is
negligible with respect to the size of the box, the integration over
the center of mass can be approximated by $V$.

In the following we shall substitute Eq.~(\ref{kcycle3}) into Eq.~(\ref{Zbec2}),
assuming that the contribution from permutations having very large cycles
is suppressed: that coincides with the approximation made in Ref.~\cite{elser}. 
The approximation is expected to work
well, apart from very close to the condensation temperature. 
Data reported in Table~\ref{tabwrap} give us the opportunity to check directly 
if this assumption is reasonable for our ensemble of monopole trajectories. We see
that indeed large values of $k$ are exponentially suppressed, so that only cycles 
with small $k$ can be found, apart from very close to $T_c$. Therefore we shall assume
the validity of Eq.~(\ref{kcycle3}), being aware that violations are expected close
to $T_c$: such violations are part of the 
finite size effects arising close to a critical point.

With the assumption above, the partition function in Eq.~(\ref{Zbec2}) becomes
\beq
Z = \frac{1}{N!} \sum_P \prod_k
{\left( \frac{V}{\lambda^3} \frac{1}{k^{3/2}} \right)}^{n_k} = 
\sum_{n_1,n_2\ldots n_N} \frac{1}{N!} C(n_1,n_2\ldots n_N)
\prod_k {\left( \frac{V}{\lambda^3} \frac{1}{k^{3/2}} \right)}^{n_k}
\label{ZHeintegrated}
\eeq
where we made explicit the partition into cycles 
($n_1$ $1$-cycles, $n_2$ $2$-cycles, \ldots $n_k$ $k$-cycles)
of each permutation $P$. The factor $C(n_1,n_2\ldots n_N)$ accounts
for the multiplicity of each partition
\mbox{$\{n_1,n_2\ldots n_N\}$}:
\beq
C(n_1,n_2\ldots n_N) = \frac{N!}{1^{n_1} 2^{n_2}\ldots n_1! n_2! \ldots} \, ,
\label{cfact}
\eeq
as it follows from a simple calculation.

The exact computation of the sum in \rep{ZHeintegrated} is not
simple, due to the presence of the constraint $\sum_k k n_k = N$ on the
possible partitions, which must sum up to the total number of
particles. This difficulty can be avoided by relaxing the constraint, i.e. by
switching to the Grand Canonical formalism.
The Grand Canonical partition function, ${\cal Z} \equiv {\rm Tr}(e^{-(H - \mu N)/T})$, can be easily 
computed starting from Eqs.~(\ref{ZHeintegrated}) and (\ref{cfact}):
\beq
{\cal Z} = \sum_{n_1,n_2\ldots } \frac{\exp\left( \frac{\mu}{T} \sum_k k n_k \right)}{1^{n_1} 2^{n_2}\ldots n_1! n_2! \ldots}
\prod_k {\left( \frac{V}{\lambda^3} \frac{1}{k^{3/2}} \right)}^{n_k}
= \prod_k \exp \left( \frac{V e^{\mu k /T}}{\lambda^3 k^{5/2}} \right)
\label{ZGCbec}
\eeq
and from the Grand Canonical partition function the average density of $k$-cy\-cles easily follows:
\beq
\rho_k \equiv \frac{\langle n_k \rangle}{V} = \frac{e^{ - \hat\mu k}}{\lambda^3\ k^{5/2}}
\label{rhok}
\eeq
where we have defined the dimensionless chemical potential $\hat\mu \equiv - \mu/T$, which 
satisfies the constraint $\hat\mu \geq 0$ (i.e. $\mu \leq 0$).

The expression for the total particle density is then 
\beq
\frac{N}{V} \simeq \rho =  \sum_{k = 1}^{\infty} k \rho_k = \frac{1}{\lambda^3} \sum_{k = 1}^{\infty}  
\frac{e^{ - \hat\mu k}}{k^{3/2}} \, .
\label{totdensity}
\eeq
Of course last expression must be equal to the usual result for the density of an ideal non-relativistic boson
gas which is obtained when working in momentum space: that can be easily verified by recalling one 
of the integral expressions for the polylogarithm ${\rm Li}_{s}(z) = \sum_k z^k/k^s$, 
in our case in particular:
\beq
\sum_k \frac{e^{-\hat\mu k}}{k^{3/2}} = \frac{2}{\sqrt{\pi}} \int_0^\infty dx \frac{\sqrt{x}}{e^{\hat\mu} e^x - 1} \, .\label{polylog}
\eeq
The expression for the total density is bounded by an upper limit which is reached for $\hat\mu = 0$
\beq
\rho_{\rm max} = \frac{1}{\lambda^3} \sum_{k = 1}^{\infty}  
\frac{1}{k^{3/2}} \simeq \frac{2.612}{\lambda^3} \, .
\eeq
For larger densities or, at fixed density, for lower temperatures (larger $\lambda$) our description
is inadequate and it is necessary to allow for macroscopic cycles or equivalently, when working in momentum space,
for a macroscopic number of particles occupying the ground (zero momentum) state: that corresponds
to Bose-Einstein condensation. The critical temperature (density) is signalled by the vanishing of the chemical potential.

If we had a set of trajectories in configuration space sampled from the path integral 
representation of an ideal non-relativistic Bose-Einstein gas at various different temperatures,
we could measure the distribution of $k$-cycles, i.e. the densities $\rho_k$, 
and then, by fitting the expected dependence given in Eq.~(\ref{rhok}), 
we could determine the chemical potential $\hat\mu$ as a function of $T$, in order to find numerically
the critical temperature $T_{\rm BEC}$ at which $\hat\mu$ vanishes and Bose-Einstein condensation begins.

The advantage of treating the condensation of the ideal gas in configuration space
rather than in momentum space should be obvious at this stage: we have a direct connection
with the information about the monopole ensemble which can be retrieved by lattice
QCD simulations. 
The aim of our analysis is indeed to try a similar thing with our ensemble of monopole trajectories, i.e. to
determine if some critical temperature $T_{\rm BEC}$
exists at which the analogous of the chemical potential vanishes, and what is its
relation with the deconfinement temperature $T_c$. There are of course 
various caveats related to particle-particle interactions, 
that we shall discuss in the next subsection.

\begin{figure}
\begin{center}
\includegraphics*[width=0.6\textwidth]{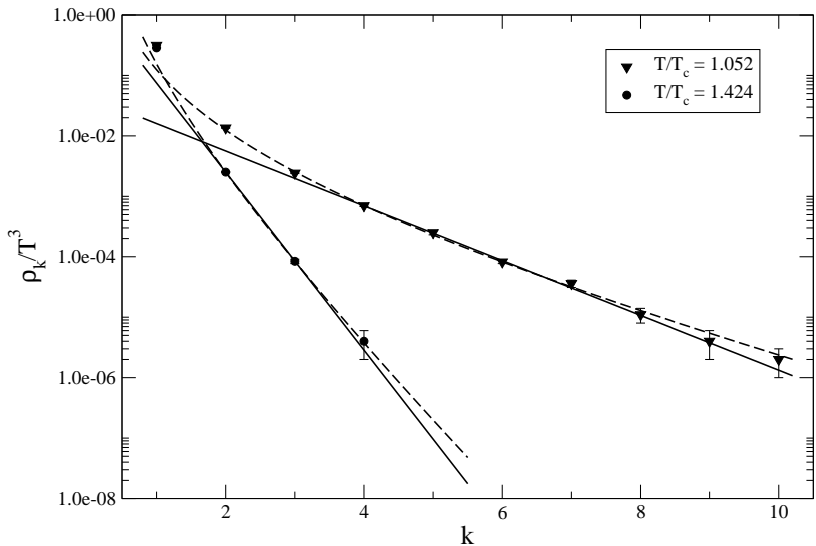}
\caption{
Fit of the densities $\rho_k$ according to $e^{-\hat\mu k}/k^{5/2}$ 
(dashed line) and according to $e^{-\hat\mu k}$ (solid line) for two 
values of the temperature. 
}
\label{fitrhok}
\end{center}
\end{figure}

\begin{table}
\begin{center}
\begin{tabular}{|c|c|c|c|c|}
\hline
$T/T_c$ & $\mu(\alpha = 3)$ & $\mu(\alpha = 2.5)$ & $\mu(\alpha = 2)$ & $\mu(\alpha = 0)$  \\
\hline $1.017^a$ & 0.43(4)  & 0.51(3)  & 0.61(4)  & 0.98(4) \\
\hline $1.052^b$ & 0.49(5)  & 0.58(4)  & 0.68(3)  & 1.00(4) \\
\hline $1.095^a$ & 0.75(5)  & 0.86(4)  & 1.02(4)  & 1.46(4) \\
\hline $1.168^b$ & 1.16(4)  & 1.32(4)  & 1.44(5)  & 1.96(6)  \\
\hline $1.187^a$ & 1.11(6)  & 1.29(5)  & 1.36(6)  & 1.89(6) \\
\hline $1.295^a$ & 1.67(6)  & 1.85(9)  & 2.03(9)  & 2.70(15)\\
\hline $1.315^b$ & 1.89(4)  & 2.04(6)  & 2.20(8)  & 2.85(15)\\
\hline $1.424^a$ & 2.18(6)  & 2.38(6) & 2.58(6) & 3.4(1)  \\
\hline $1.503^b$ & 2.64(10) & 2.8(1)   & 3.0(1)   & 3.8(1)  \\
\hline $1.582^a$ & 2.69(10) & 2.9(1)   & 3.1(1)   & 3.92(8) \\
\hline $1.754^b$ & 3.16(8)  & 3.37(10) & 3.57(10) & 4.37(15)\\
\hline
\end{tabular}
\end{center}
\caption{Chemical potentials obtained for different temperatures by fitting the densities
$\rho_k$ according to $\rho_k \propto e^{-\hat\mu k}/k^\alpha$, for different values of $\alpha$.
}
\label{tabchem}
\end{table}

\subsection{BEC of thermal monopoles}
\label{monbec}

The ensemble of thermal Abelian monopoles that we are investigating is surely different from an ideal 
gas of non-relativistic identical particles. 
The analysis of Ref.~\cite{monden} indeed has shown the presence of particle-particle interactions,
which are attractive in the monopole-antimonopole case and repulsive in the monopole-monopole
case. 
The presence of those interactions interferes with many steps of the above derivation
for the densities $\rho_k$, for instance it is not possible
to rewrite the contribution of a single $k$-cycle as in Eq.~(\ref{kcycle2}).

One of the effects of interactions is visible in the particle density around 
a given thermal monopole, which is suppressed (enhanced) for monopoles (antimonopoles) with
respect to the average density. In the monopole-monopole case 
the suppression extends, close to $T_c$, 
over distances of the order of 1 fm: as we shall discuss
in the next Section, such distance is larger than the typical thermal wavelength $\lambda$
which can be measured on trajectories wrapping only one time. 
As a consequence, interactions should clearly affect 
monopole trajectories corresponding to $k$-cycles with $k > 1$;
the typical size of these loops should be $\lambda \sqrt{k}$ in the non-interacting case,
but is expected to be larger due to monopole-monopole repulsion:
that corresponds indeed to the outcome of the analysis to be presented in the next 
Section.

Apart from the interactions among particles belonging to the same loop, one should also 
take into account loop-loop interactions, which can  be both repulsive or attractive 
(a monopole $k$-cycle will repel (attract) a monopole (antimonopole) $k'$-cycle),
so that it is also not possible to factorize the contribution of each permutation into 
the contribution of different $k$-cycles, as in Eq.~(\ref{Zbec2}).

Even if we have some partial information about 
interactions from the analysis of Ref.~\cite{monden}, 
taking them properly into account, in order to repeat the description of BEC
along the same lines reported above for the free case, is not an easy task.

Instead of looking for an exact solution, let us try to understand which similarities
and differences should be expected for the dependence of the density $\rho_k$ on $k$,
with respect to the free case reported in Eq.~(\ref{rhok}). On general grounds one may 
expect some finite free energy cost needed to add one particle to a $k$-cycle, 
i.e. to go from $k$ to $k+1$, playing the role of an effective chemical potential,
plus some interaction dependent contribution, so that:
\beq
\rho_k = e^{ - \hat\mu k} f(k)
\label{rhokint}
\eeq
where $f(k)$ is some function decreasing less than exponentially with $k$.
We shall refer to $\hat \mu$ as chemical potential in the following, however it should
be clear that in the present context, in which we are dealing with a neutral plasma
of monopoles and antimonopoles, it should be better regarded as a parameter for an effective 
description of the distribution in the number of $k$-cycles.

The result obtained in the free case, $f(k) = (\lambda^3 k^{5/2})^{-1}$, can be modified
by interactions in various ways. In the case of a dilute hard sphere gas model, reported
in Ref.~\cite{elser}, $f(k)$ is the same as in the free case plus corrections
of order $r/\lambda$, where $r$ is the sphere radius. Part of the interactions
could be also taken into account by an effective dynamical mass, as done by Feynman
for the study of ${}^4$He. In general we may expect a leading contribution 
$f(k) \propto 1/k^\alpha$, where $\alpha$ could be different
from $5/2$ (this is also the way in which relativistic
effects should show up). In any case, the approach to condensation should be signalled by the 
vanishing of the chemical potential, i.e. at the condensation point large $k$-cycles should 
cease to be suppressed exponentially in $k$. This is exactly what we want to check on 
our data reported in Table~\ref{tabwrap}. As we shall discuss soon, the outcome is that
the vanishing of the chemical potential happens at a point which is compatible with
$T_c$ and that this result is remarkably stable for various different choices of the 
function $f(k)$.

\begin{figure}
\begin{center}
\includegraphics*[width=0.6\textwidth]{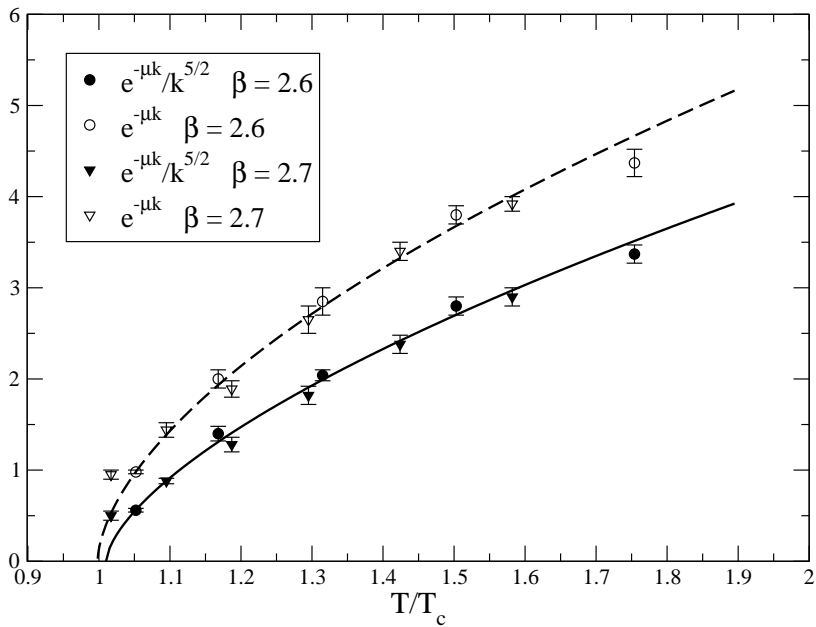}
\caption{
Chemical potentials reported in Table~\ref{tabchem} for $\alpha = 0$ and $\alpha = 2.5$ and two 
different lattice spacings, together with a fit of them according to Eq.~(\ref{critchem}).
}
\label{figcritchem}
\vspace{1cm}
\end{center}
\end{figure}

\begin{table}
\begin{center}
\begin{tabular}{|c|c|c|c|}
\hline
$\alpha$ & $T_{\rm BEC}/T_c$ & $\nu'$ & $\chi^2/{\rm d.o.f.}$ \\ 
\hline 3 & 1.005(13)   & 0.71(5) &  2.24 \\
\hline 2.5 & 1.000(12)   & 0.68(5) &  1.23 \\
\hline 2 & 0.989(13)  & 0.68(5) &  1.72 \\
\hline 0 & 0.988(15)  & 0.61(5) & 2.34 \\
\hline
\end{tabular}
\end{center}
\caption{Results of the fit of chemical potentials according to 
$\hat\mu = A\ (T - T_{\rm BEC})^{\nu'}$, for various different values of $\alpha$. The lowest
temperature, $T = 1.017$, has always been discarded from the fit.
}
\label{tabfit}
\end{table}

We have tried to fit our data for the normalized densities $\rho_k/T^3$ according to
$ A e^{-\hat\mu k}/k^\alpha$. In order to obtain reasonable values for
$\chi^2/{\rm d.o.f.}$ we had to take into account only data with $k > 2-3$ for
$T \leq 1.2\ T_c$ and with $k > 1$ for higher temperatures. If the $\alpha$ parameter
is left free, the $\chi^2$ is minimized for $\alpha$ around 2 for most of the explored
temperatures, however reasonable fits are obtained for a larger range 
of $\alpha$ values\footnote{Notice that a 
value $\alpha > 2$ is to be preferred for the asymptotic large $k$ 
behavior, since in this case the total number of particles
$\sum k \rho_k$ is convergent even for $\hat\mu = 0$, thus claiming for
a macroscopic number of particles occupying the ground state.}.
A few examples of such fits are shown in Fig.~\ref{fitrhok}, while in
Table~\ref{tabchem} we report results obtained for the chemical potential
for various values of $\alpha$ going from 0 to 3.

As a second step, we have tried to fit the values obtained for the chemical
potential according to a critical behavior:
\beq
\hat\mu = A\ (T - T_{\rm BEC})^{\nu'}
\label{critchem}
\eeq
which predicts $\hat\mu$ to vanish, as the condensation temperature $T_{\rm BEC}$ 
is approached from above, with a critical exponent $\nu'$. 
Such functional dependence describes reasonably well data reported in Table~\ref{tabchem},
for each different value of $\alpha$, if the lowest temperature, $T = 1.017\ T_c$, is discarded: 
this is expected, since the vanishing of $\hat\mu$ implies the appearance 
of $k$-cycles with arbitrarily large $k$, i.e. cycles of permutating particles with arbitrarily 
large spatial extension, corresponding to a diverging correlation length; 
this is not possible on a finite lattice hence the dependence Eq.~(\ref{critchem}) must be valid 
not too close to the condensation temperature. One might wonder whether such finite size effects
are also present, on the lattices used for our investigation, 
for higher values of $T$: this issue is discussed in detail in Section~\ref{finitesize},
where we show that this is not the case already for $T = 1.052\ T_c$.

We report in Table~\ref{tabfit} the results for $T_{\rm BEC}$
and $\nu'$ obtained from fits limited to a temperature 
range $1.052 \leq T/T_c \leq 1.582$; values do not change within
errors if the range is slightly changed (but not including the lowest temperature
$T = 1.017\ T_c$). A few examples of such fits are also reported in Fig.~\ref{figcritchem}.
The values obtained for $\chi^2/{\rm d.o.f.}$ are reasonable, taking into account
that we have mixed data obtained for different values of the lattice spacing 
and that scaling violations, even if small, may be present; 
the ansatz given in Eq.~(\ref{critchem}) seems to work slightly better for the 
chemical potentials obtained in the case $\alpha = 2.5$, i.e. the value expected for
free bosons.

As can be inferred from Table~\ref{tabfit}, the dependence of $T_{\rm BEC}$ and
$\nu'$ on the exponent $\alpha$, used to deduce the chemical potentials from
the densities $\rho_k$, is very mild or even negligible within errors.
That makes the main result of our analysis stronger and less dependent on the way
we deal with interactions: $T_{\rm BEC}$ is compatible within errors with the
critical temperature $T_c$ at which the SU(2) pure gauge theory deconfines.
Therefore we have proved that, as we proceed from higher to lower temperatures, the onset of confinement
is associated with the condensation of thermal monopoles present in the 
deconfined phase, as expected in the dual superconductor scenario for color confinement.

If monopole condensation indeed coincides with deconfinement, a second important
question regards the relation of the critical exponent $\nu'$ with
the critical exponents of the second order SU(2) deconfinement transition, which
belongs to the 3d Ising universality class. As 
clearly visible from Table~\ref{tabfit}, the value obtained for $\nu'$ is 
compatible or slightly larger than the critical exponent related to the diverging behavior
of the correlation length at the transition, $\xi \sim (T - T_c)^{-\nu}$,
which for Ising 3d is $\nu \sim 0.63$: in the following we shall try to give
a possible interpretation of this fact.

The increase in the correlation length $\xi$, observed when approaching the critical
condensation temperature from above, may be related to the appearance of larger and larger
$k$-cycles, corresponding to groups of permutating particles which extend over
larger and larger spatial extensions. We may assume that $\xi$ grows proportionally
to the typical spatial extension of $k$-cycles.
For a distribution of $k$-cycles
$\rho_k \sim e^{-\hat\mu k}/k^\alpha$, the average value of $k$ is $\langle k \rangle \propto 1/\hat\mu$;
therefore, if we knew how the typical spatial extension of a $k$-cycle depends on $k$, the dependence of 
$\xi$ on $\hat\mu$ would be easily obtained. In the free boson case, the typical
spatial extension of a $k$-cycle is $\propto \sqrt{k}$, corresponding to a random walk behavior.
However the size is expected to grow faster with $k$ because of particle-particle repulsive interactions: 
as a limiting case, opposite to the random walk behavior, the tyical size of a $k$-cycle may grow linearly 
with $k$ if interactions induce ordered linear structures 
in the set of permutating particles (e.g. the particles are typically disposed 
on some closed smooth line). In general
we expect $\xi \propto 1/\hat\mu^\omega$, where we have introduced the exponent $\omega$ which dictates
how the size of $k$-cycles grows and is expected to be in 
the range $1/2 \leq \omega \leq 1$: lower values of $\omega$ could be possible if the permutating
particles are more closely packed than in the non-interacting case, but this is not expected in presence of repulsive
interactions. 
Since $\xi \sim (T - T_c)^{-\nu}$, we expect 
$\hat\mu \sim (T - T_c)^{\nu/\omega}$, hence $\nu' = \nu/\omega$.

From Table~\ref{tabfit} we conclude that $\nu'$ is compatible with 
$\nu$, as expected if $\omega \sim 1$: this suggests
that the typical sets of permutating thermal monopoles contributing to the path integral
form linearly ordered structures, which are likely induced by interactions. Therefore
the typical cluster of monopole currents corresponding to a $k$-cycle contribution 
should lie on a sort of time oriented closed surface (see Ref.~\cite{chernodub09}
for a recent discussion about the possible appearance of similar structure right above 
$T_c$).

\begin{table}
\begin{center}
\begin{tabular}{|c|c|c|c|c|c|}
\hline
  &  $L_s = 32$  &  $L_s = 40$  &  $L_s = 48$  &  $L_s = 56$  &  $L_s = 64$  \\
\hline
\hline $\rho_1/T^3$ & 0.317(4) & 0.316(4) & 0.315(5) & 0.315(4) & 0.315(5) \\
\hline $\rho_2/T^3$ & $1.45(2)\ 10^{-2}$& $1.37(2)\ 10^{-2}$& $1.35(1)\ 10^{-2}$& $1.36(2)\ 10^{-2}$& $1.33(1)\ 10^{-2}$
\\
\hline $\rho_3/T^3$ & $2.51(7)\ 10^{-3}$& $2.41(5)\ 10^{-3}$& $2.42(4)\ 10^{-3}$& $2.38(6)\ 10^{-3}$& $2.39(4)\ 10^{-3}$
\\
\hline $\rho_4/T^3$ & $5.0(3)\ 10^{-4}$   & $6.1(3)\ 10^{-4}$   & $7.0(2)\ 10^{-4}$   & $6.6(3)\ 10^{-4}$   & $6.7(2)\ 10^{-4}$   
 \\
\hline $\rho_5/T^3$ & $0.94(15)\ 10^{-4}$ & $1.8(2)\ 10^{-4}$ & $2.5(2)\ 10^{-4}$ & $2.3(2)\ 10^{-4}$ & $2.30(15)\ 10^{-4}$ 
 \\
\hline $\rho_6/T^3$ & $0.13(5)\ 10^{-4}$ & $0.52(8)\ 10^{-4}$ & $0.82(5)\ 10^{-4}$ & $0.80(13)\ 10^{-4}$ & $0.88(7)\ 10^{-4}$ 
\\
\hline $\rho_7/T^3$ & $0.2(2)\ 10^{-5}$   & $1.4(5)\ 10^{-5}$   & $3.6(5)\ 10^{-5}$   & $3.0(7)\ 10^{-5}$   & $2.7(5)\ 10^{-5}$   
\\
\hline $\rho_8/T^3$ & -- & $4(3)\ 10^{-6}$ & $11(3)\ 10^{-6}$ & $14(4)\ 10^{-6}$ & $11(3)\ 10^{-6}$ 
\\
\hline $\rho_9/T^3$ & -- &  $1(1)\ 10^{-6}$ & $4(2)\ 10^{-6}$ & $3(2)\ 10^{-6}$ & $6(2)\ 10^{-6}$ 
\\
\hline $\rho_{10}/T^3$ & -- &  $1(1)\ 10^{-6}$ & $2(1)\ 10^{-6}$ & $2(1)\ 10^{-6}$ & $3(1)\ 10^{-6}$ 
\\
\hline
\hline $\mu(\alpha = 2.5)$ & 1.25(6) & 0.76(4) & 0.58(4) &  0.57(3) & 0.56(3) \\
\hline 
\end{tabular}
\end{center}
\caption{Normalized densities for $T = 1.052\ T_c$ ($\beta = 2.60$ and $N_t = 10$) and 
various spatial lattices $L_s$. The chemical potentials obtained for
$\alpha = 2.5$ are also reported. For the explorable number of
wrappings, finite size effects are negligible, within errors, for $L_s \geq 48$.
}
\label{tabwrapfsize}
\end{table}

\begin{figure}
\begin{center}
\includegraphics*[width=0.6\textwidth]{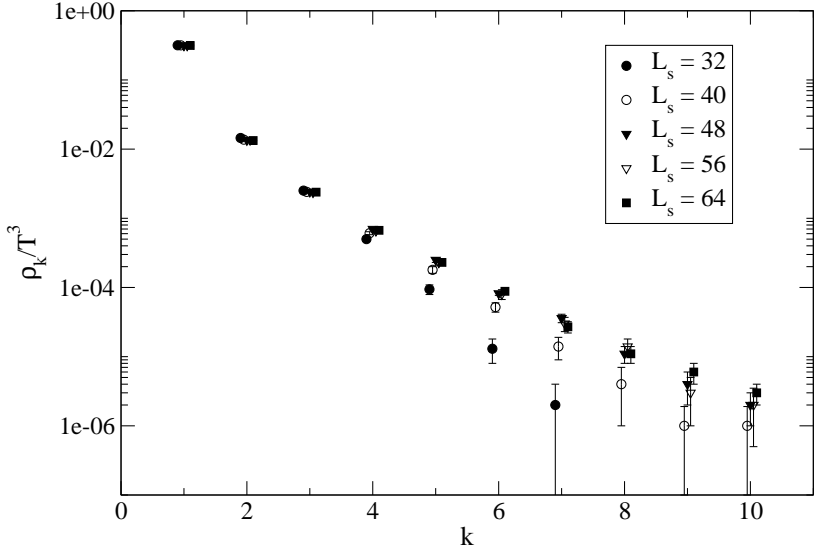}
\caption{
Dependence of the normalized densities $\rho_k/T^3$ on the 
spatial lattice size for $\beta = 2.60$ and $N_t = 10$ ($T/T_c = 1.052$).
}
\label{figfsize}
\vspace{1cm}
\end{center}
\end{figure}

\subsection{Finite size effects}
\label{finitesize}

As discussed above, it is natural to expect that the density 
of trajectories with a high wrapping number may be affected by the 
finiteness of the system in the spatial directions. That may be a source
of systematic error in our analysis, in particular for the temperatures
closest to the transition, where the density of multiple wrapping trajectories
is higher. In fact, if the typical cycle extension is really linked to the correlation
length of the system, such effects are another way of looking at the saturation
of $\xi$ as the critical temperature is approached on a finite lattice.
Our data have been obtained on two different combinations
of lattice size and bare coupling, i.e. $L_s = 48$ at $\beta = 2.6$ and 
$L_s = 64$ at $\beta = 2.7$: in both cases the physical spatial size
is approximately the same, $T_c L_s a(\beta) \sim 4.5$. 

Therefore, in order to check for finite size effects, we have considered
the lowest temperature included in our analysis, i.e. $T = 1.052\ T_c$ 
(corresponding in our numerical setup to $\beta = 2.6$ and $N_t = 10$),
where we have repeated the determination of the densities of multiple wrapping trajectories and of 
the chemical potential $\hat \mu$ on various different spatial lattice sizes,
$L_s = 32,40,48,56,64$, corresponding to $T_c L_s a(\beta)$ ranging from 3 to 6.

Results obtained for the densities of multiple wrapping trajectories are reported in 
Table~\ref{tabwrapfsize} and in Fig.~\ref{figfsize}. In Table~\ref{tabwrapfsize} we also
report the determination of the chemical potential $\hat \mu$ in the free boson 
approximation ($\alpha = 2.5$) on the different lattice sizes. 

Finite size effects are clearly visible on the smallest lattice and result in a stronger suppression
of trajectories with a large number of wrappings, or equivalently in a higher
value of $\hat \mu$. 
However results are compatible within errors for the three largest lattices, $L_s = 48,56,64$.
We take that as evidence that finite size systematic effects are negligible, within our 
current statistical uncertainties, for spatial lattice sizes $a L_s \geq 4.5\ T_C^{-1}$ 
and for temperatures $T \geq 1.052\ T_c$. Such systematic effects therefore do not affect
our analysis presented above.

\section{Monopole mass}
\label{section4}

We have followed different strategies in order to determine a temperature
dependent effective monopole mass. As we shall discuss, different definitions lead
to different results, so that no unambiguous determination can be achieved; nevertheless
one can identify some common features.

\begin{table}
\begin{center}
\begin{tabular}{|c|c|c|}
\hline
$T/T_c$ &  $m_{\rm BEC}/T_c$ & $m_{\rm FLUC}/T_c$  \\
\hline $1.017^a$ & 2.93(25)  & 1.56(1)    \\
\hline $1.052^b$ & 2.41(22)  & 1.71(1)    \\
\hline $1.095^a$ & 3.67(19)  & 2.15(1)    \\
\hline $1.168^b$ & 5.22(35)  & 2.62(1)    \\
\hline $1.187^a$ & 5.28(29)  & 3.00(2)    \\
\hline $1.295^a$ & 8.72(63)  & 4.18(2)    \\
\hline $1.315^b$ & 9.84(76)  & 3.93(1)    \\
\hline $1.424^a$ & 12.6(8)   & 5.63(2)    \\
\hline $1.503^b$ & 19(2)     & 5.53(1)    \\
\hline $1.582^a$ & 18(2)     & 7.29(2)    \\
\hline $1.754^b$ & 25(3)     & 7.36(1)    \\
\hline
\end{tabular}
\end{center}
\caption{Monopole masses, determined for different temperatures according to the definitions
given in Eqs.~(\ref{mass1}) and (\ref{mass2}).
}
\label{tabmass}
\end{table}

\begin{figure}
\begin{center}
\includegraphics*[width=0.6\textwidth]{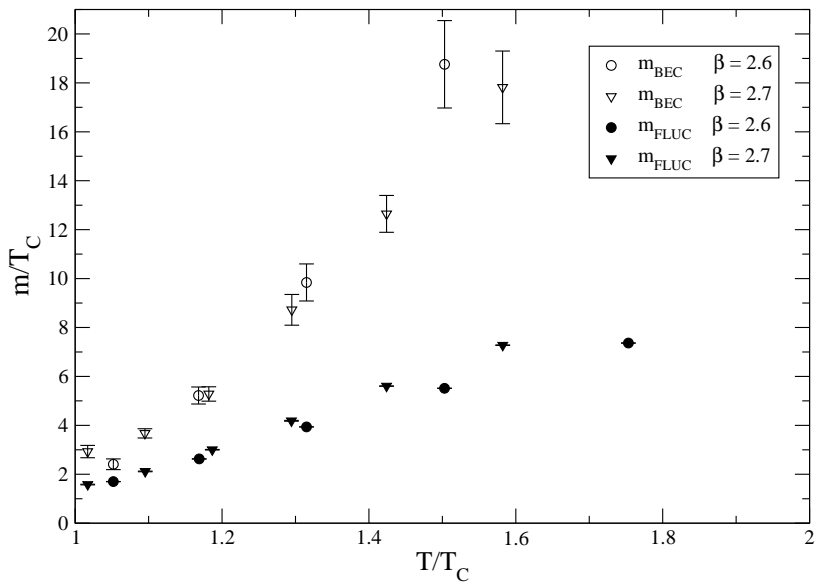}
\caption{
Monopole masses as a function of $T/T_c$, as reported in Table~\ref{tabmass}.
}
\label{figmass}
\vspace{1cm}
\end{center}
\end{figure}

The first strategy is to consider the fit of the densities of wrapping trajectories
$\rho_k$ according to the non-relativistic free boson prediction, Eq.~(\ref{rhok}), 
and extract from the coefficient $1/\lambda^3$ an estimate for an effective dynamical
mass 
\beq
m_{\rm BEC} = \frac{2 \pi}{T \lambda^2} \label{mass1} \, .
\eeq
If we consider the same fits used to compute the chemical potentials reported in
Table~\ref{tabchem} for $\alpha = 2.5$, we obtain the masses reported in the second column of 
Table~\ref{tabmass} and also in Fig.~\ref{figmass}. The dependence on the lattice
spacing seems negligible within errors.
If we neglect the determination obtained for $T = 1.017\ T_c$, where as discussed in previous 
Section finite size effects could be significant, we see a clear trend for an effective
monopole mass rapidly decreasing from values of the order of $20\ T_c$ for $T \sim 1.5\ T_c$
down to $m_{\rm BEC} \sim T_c$ at $T_c$. Of course when $m_{\rm BEC}$ is of the order
of $T_c$ the non-relativistic approximation breaks down, however even a linear fit taking
into account only values for $T \geq 1.095\ T_c$, where $m_{\rm BEC}$ is already
a few times $T_c$ and relativitic corrections can be neglected, leads to predict
$m_{\rm BEC} \sim T_c$ at the transition.

A different strategy is to relate the monopole mass to 
the spatial fluctuations of monopole trajectories: those should be more and more damped
as the mass increases.
Consider again the path integral of a non-relativistic particle of mass $m$
and the average squared spatial fluctuation of the particle periodic path, defined as
\beq
\Delta x^2 \equiv T \int_0^{1/T}dt  \langle (\vec x(t)-\vec
x(0))^2 \rangle\, ,
\label{dx2}
\eeq
where the average is taken over the path integral distribution. That is directly
related to the mass of the particle. A simple calculation
shows that 
for a free non-relativistic particle one has $\Delta x^2 = 1/(2 m T)$, while 
for a particle in a harmonic potential with elastic constant $m \omega^2$ one obtains:
$\Delta x^2 = -{6}/({\beta m \omega^2}) + {3 \coth(\omega\beta/2)}/{(m \omega)}$.

\begin{table}
\begin{center}
\begin{tabular}{|c|c|c|c|}
\hline
$T/T_c$ &  $T^2\ \Delta x^2_{k = 1}$ &  $T^2\ \Delta x^2_{k = 2}$ &  $T^2\ \Delta x^2_{k = 3}$ \\
\hline $1.017^a$ & 0.322(6)      & 1.85(3)  & 3.16(6)  \\
\hline $1.052^b$ & 0.309(2)      & 1.73(2)  & 3.10(3)  \\
\hline $1.095^a$ & 0.259(4)      & 1.46(3)  & 2.71(7)  \\
\hline $1.168^b$ & 0.222(1)      & 1.23(2)  & 2.42(5)  \\
\hline $1.187^a$ & 0.198(1)      & 1.58(2)  & 2.06(6)  \\
\hline $1.295^a$ & 0.155(1)      & 0.92(2)  & 1.90(13)  \\
\hline $1.315^b$ & 0.1670(7)     & 0.915(12)  & 1.80(6)  \\
\hline $1.424^a$ & 0.1271(5)     & 0.741(8)  &  1.50(7) \\
\hline $1.503^b$ & 0.1362(4)     & 0.722(6)  &  1.45(6) \\
\hline $1.582^a$ & 0.1087(3)     & 0.656(9)  &  1.23(7) \\
\hline $1.754^b$ & 0.1109(3)     & 0.616(8)  &  1.30(8) \\
\hline
\end{tabular}
\end{center}
\caption{Average squared fluctuations measured as reported in Eq.~(\ref{ldx2}),
 as a function of $T/T_c$, for trajectories having
wrapping numbers $k = 1,2,3$.
}
\label{tabfluc}
\end{table}

On the lattice the mean
squared monopole fluctuation is defined as 
\beq
a^{-2} \Delta x^2 = 
\frac{1}{L}\sum_{i=1}^L
d_i^2 \, ,
\label{ldx2}
\eeq 
where $d_i$ is the squared spatial
distance (accounting for periodical boundary conditions)
between the starting site of the monopole current at $t=0$
and the current position after $i$ steps along the monopole trajectory, while $L$ is the total length
of the trajectory (all quantities measured in lattice spacing units). We have measured this quantity 
on trajectories wrapping only one time and tried to relate it to the monopole
mass. 
We have considered the free particle approximation, thus definining a mass
\beq
m_{\rm FLUC} = \frac{1}{2 T \Delta x^2} \, ,
\label{mass2}
\eeq
and
obtaining the results reported in Table~\ref{tabmass} and in Fig.~\ref{figmass}.
A better approximation would be to consider interactions at a mean field level. One can
take the charge distribution around a monopole from the correlation
function $g(r)$ computed in Ref.~\cite{monden} and determine the potential 
around the position of a monopole induced by nearby monopoles and 
antimonopoles, then approximating it by an attractive harmonic potential: we have tried
this procedure obtaining negligible corrections to the free particle approximation.

The mass obtained in this way shows some scaling deviations, especially at high $T$,
which can be explained as follows. Wrapping monopole
trajectories are not strict one-dimensional lines as in the one-particle path integral,
but have some finite thickness, part of which is due to unphysical ultraviolet
fluctuations in the form of monopole-antimonopole loops at the scale of the UV cutoff,
which may be attached to a physical wrapping trajectory. Such unphysical fluctuations add
to the physical ones that we are investigating. 
It is not easy to 
give a definition for a proper subtraction of the unphysical contribution.
However we notice that it becomes negligible close to $T_c$, where physical
fluctuations of the trajectory become important: that is signalled by the fact
that scaling is better and better as we approach $T_c$.

The value of the monopole mass obtained from spatial fluctuations of trajectories 
differs significantly from that obtained in the BEC approach:
that is expected since $m_{\rm BEC}$ is related to the properties of multiple
wrapping trajectories, while $m_{\rm FLUC}$ to those of single wrapping trajectories:
in both cases we have relied on some effective free particle behavior, but
consistent results would have been obtained only if the particles were really free.
It is however remarkable that also $m_{\rm FLUC}$ decreases rapidly as the transition
temperature is approached, getting compatible with $m_{\rm BEC}$ if extrapolated down
to $T_c$ (see Fig.~\ref{figmass}).

Finally let us comment on $\Delta x^2$ measured on trajectories wrapping more than
one time. In Table~\ref{tabfluc} we report the values of $T^2\ \Delta x^2 (k)$ measured
for $k = 1,2,3$ at various temperatures.
In the case of free particles one would expect $\Delta x^2 (k) \sim k \Delta x^2 (k = 1)$,
instead, as anticipated in Section~\ref{monbec}, values increase faster with $k$
because of monopole-monopole repulsions. In particular $\Delta x^2 (k = 2)$
and $\Delta x^2 (k = 3)$ are respectively about 5-6 times and 10 times  larger than
$\Delta x^2 (k = 1)$.

\section{Discussion and Conclusions}
\label{section5}

Let us summarize the main results obtained in this work. We have considered the statistical
distribution in the number of wrappings in the Euclidean time direction, for 
thermal monopole trajectories exposed after MAG Abelian projection in the deconfined phase
of SU(2) pure gauge theory. We have shown that such distribution is compatible with that
of a Bose gas condensing exactly at the confinement/deconfinement critical temperature $T_c$
of the pure gauge theory. This result represents a strong link between confinement, monopole condensation
and thermal monopoles.
We have also found  that in the approach to the criticality in the monopole
ensemble,
as $T$ decrease to $T_c$ from above,
 the monopole masses are significantly reduced,  and their 
interactions weaken.
That fits into the overall idea of plasma gradually switching from electric to
magnetic one.
Our quantitative results about cluster densities shed further light on
properties
and interactions of the objects we call monopoles.
Our investigation leaves some aspects that should be better understood and clarified
and that we partially discuss in the following.

1) We still do not understand the exact physical nature of thermal monopoles. The definition
of monopoles studied in this work is strictly linked to a particular choice of Abelian projection, namely to 
that performed in Maximal Abelian Gauge, so that one should understand why some other Abelian projections 
give different, unphysical results and what non-Abelian gauge configurations are associated  
with the MAG Abelian monopoles. Some steps in this direction has been done in Ref.~\cite{dyons},
where it has been shown a clear correlation of the non-Abelian gauge action density with 
the locations of thermal monopole trajectories. 
Objects identified by MAG projection have magnetic charge, yet various other quantum numbers are 
in principle possible. They can in principle be just monopoles, or dyons, 
or (in simulations with several quark flavors) they may also have various flavors, 
due to binding to quark-antiquark pairs or  single quarks, or be
a complicated mixture of all of them. Examples of fermionic 
excitations possessing magnetic charge and spin 1/2 are well known, and in fact
required in any supersymmetric theory,  complementing all monopole multiplets. 
Seiberg-Witten ${\cal N}=2$ superconformal theories with matter have not only magnetic monopoles 
with certain flavor allocations, but they also have intricate structure of dualities, relating several different
(mutually dual) Lagrangians with the same theory. So, in simulations with quarks, one would need to separate  
Bose and Fermi-like behavior of various magnetic objects, which is by no means trivial. 
Those studies, as well as other issues related with the quest for understanding of the precise 
spectroscopy of the magnetic sector, will hopefully be continued elsewhere.

2) While the statistical distribution of wrappings strongly suggests the presence of a BEC-like phenomenon
associated with confinement, a clear description of the physical properties of the system has not 
yet been achieved. The ensemble of thermal monopoles is strongly interacting and its properties 
resemble those of a liquid~\cite{monden,liquid}, therefore one should not try to fit its properties with
those of a weakly interacting but rather with those of a strongly interacting Bose gas, like
$^4$He~\cite{cristoforetti}. The effect of interactions is clearly visible in the typical 
shape of "polymers" made up of monopoles belonging to the same $k$-cycle: we have not studied
that in detail yet, but just remarked that from the scaling of the chemical potential we predict
smooth linear shapes rather than the random walk behavior expected in the case of free particles;
further investigation on this side could clarify more universal properties of the system.

3) Another peculiarity of the system is that in fact it is a plasma of monopoles and antimonopoles
with Coulomb-like screened interactions, i.e. it resembles a 
Bose-Coulomb gas system which is neutral on average: in this sense the chemical potential that we have 
introduced in our analysis should not be regarded as a true chemical potential, but rather
as a parameter entering an effective description of the increase in the number of $k$-cycles.

4) The vanishing of the chemical potential $\hat \mu$ at the transition, with a critical 
exponent $\nu'$ (see Eq.~(\ref{critchem})) compatible with the 3d Ising critical exponent
$\nu$ of the correlation length, shows that $\hat \mu$ scales  with the same critical behaviour
of a mass. It would be interesting to better understand the meaning of this in the future,
for instance by investigating the relation between the distribution in the number of 
wrappings of monopole trajectories and other quantities of dual nature, showing a critical
behaviour above $T_c$, like the dual string tension related to the spatial 't Hooft
loop~\cite{thooftloop,khaltes,dualstring}.

5) The fact that we still do not have a consistent physical description of the system 
manifests itself, for instance, in the absence of a consistent definition of effective monopole
mass. However it is interesting to notice that different definitions tend to a common
value $m \sim T_c$
when the condensation temperature is approached, where they are also roughly in agreement 
with the estimates reported in Ref.~\cite{cristoforetti}. A similar physical scale
comes out when studying the electric-magnetic asymmetry in Landau gauge~\cite{ilgenfritz}: 
it would be important
to understand if the two scales are connected in some way.

6) The fact that we obtain a condensation temperature
$T_{\rm BEC} = T_c$, within our statistical uncertainties, by just
extrapolating data for the densities of trajectories with multiple wrappings, may be peculiar to SU(2) 
pure gauge theory, where the confinement/deconfinement transition is of second 
order nature, hence there is a critical behavior at the transition.
In different cases, like for a first order transition, hence e.g. for the SU(3) pure gauge theory,
while one expects monopole condensation to happen at $T_c$ anyway,
the statistical analysis of the distribution of wrapping trajectories above $T_c$ 
could point to a condensation temperature slightly below the actual deconfinement temperature,
i.e. to a sort of spinodal point.
In this sense $SU(2)$ pure gauge theory is a
system where the connection between confinement and monopole
condensation can be better established.
We would like to clarify this point in the future by extending our investigation to SU(3)
color gauge group, with and without dynamical quark contributions.

\acknowledgments

We thank the organizers of the 
workshop ``Non-Perturbative Methods in Strongly Coupled Gauge
Theories'' at the Galileo Galilei Institute (GGI) in Florence, where
this work has been started.
We thank D.~Antonov, C.~M.~Becchi, M.~Chernodub, M.~Cristoforetti, A.~Di~Giacomo, P.~de~Forcrand, M.~Ilgenfritz,
J.~Liao, E.~Vicari and V.~Zakharov for useful discussions.
Numerical simulations have been performed on GRID
resources provided by INFN.

\end{document}